
%
\def\squ1{g^{\mu\nu}\partial_\mu\phi\partial_\nu\phi-1}
\def\sq2{g^{\rho\sigma}\partial_\rho\phi\partial_\sigma\phi-1}
\def\ghmn{g^{\mu\nu}}
\def\glmn{g_{\mu\nu}}
\def\glrs{g_{\rho\sigma}}
\def\frac#1#2{{{#1}\over {#2}}}
\def\ghmnp#1{{g^{{#1}\mu\nu}}}
\def\glmnp#1{{g^{#1}_{\mu\nu}}}

\input harvmac

\baselineskip8pt
\Title{\vbox
{\baselineskip 6pt{\hbox{  }}{\hbox
{FIAN/TD/95-05 }}{\hbox {}}{\hbox{hep-th/9504nnn}} } }
{\vbox {\centerline{Random Surfaces in Four Dimensions}
\vskip4pt\centerline{and One-Dimensional String Theory}}}

\vskip 37 true pt
\centerline  { {Dmitri S. Dolgov\footnote {$^*$} {e-mail address:
ddolgov@lpi.ac.ru} }}

\smallskip
\smallskip
\centerline {\it I.E.Tamm Theoretical Department}
\smallskip
\centerline{\it  P.N.Lebedev Physical Institute}
\smallskip
\centerline{\it Leninsky prospekt 53, Moscow 117924, Russia}

\bigskip

\centerline {\bf Abstract}

\medskip
\baselineskip 18pt
\noindent

We consider a new action of a two-dimensional field
theory interacting with gravitational field. The action is
interpreted as the area of a surface imbedded into
four-dimensional Mincowski target space. In addition to
reparametrization invariance the new action has one extra
infinite-dimensional local symmetry with a clear geometrical
meaning.  The special gauge choice, which includes the
gauge condition of tracelessness of the energy-momentum tensor,
leads to an effective free scalar field theory. The problem of
anomalies in quantum theory and possible connection with
matrix quantum mechanics are also discussed.

\Date {April 1995}


\noblackbox
\baselineskip 16pt plus 2pt minus 2pt


\vfill\eject

\newsec{Introduction}

In recent years a lot of work was done in the field
of one-dimensional string theories, i.e. string theories with one
dimensional target space (for a review see, for example
\ref\rgins{ P.Ginsparg and  G.Moore,\  {\it Preprint
YCTP-P23-92\/} (hep-th/9304011) \semi I.Klebanov,\  {\it
Preprint PUPT-1271\/} (hep-th/9108019) \ }).  These theories are
connected with matrix models and integrable systems and that
makes possible the calculation of the correlation functions in
all orders of perturbation expansion over Riemann surfaces.
However it is difficult to generalize this connection (and to
apply the corresponding powerful techniques) for the most
physically interesting case of $D$-dimensional string theories
($D>1$).  Instead, the one-dimensional string theories one
usually consider as a toy models.

In this work we are going to argue that some
definite {\bf one-dimensional string theory},
i.e. some field theory on two-dimensional manifold
of one scalar field $\phi(\sigma,\tau)$ interacting with
(two-dimensional) metric $\glmn(\sigma,\tau)$
effectively could be
interpreted as a  string theory
in {\bf four-dimensional} Mincowski space.

The idea is, roughly speaking, to consider
the two-dimensional metric $\glmn$ which appears in the theory
as the metric, induced by an imbedding of the
two-dimensional world-sheet into {\bf three-dimensional}
Euclidean space, and to consider the only field $\phi$ as the
remaining time-coordinate $X^0$.  In other words, one can encode
the {\bf three} functions $X^i(\sigma,\tau)$ ($i=1,...,3$)
corresponding to Euclidean coordinates into new dynamical
variables $\glmn(\sigma,\tau)$ ($\mu,\nu=0,1$) as follows:  $$
\glmn=\partial_\mu X^i \partial_\nu X^i
$$

Than it is easy to rewrite in new terms  the area of
the world-sheet thus obtaining an action depending
on $\glmn$ and $\phi$,
and in quantization
in the path integral one may replace
the integration over imbeddings $X^\alpha(\sigma,\tau)$ where
$\alpha=0,...,3$ with the integration over $\glmn(\sigma,\tau)$
and $\phi(\sigma,\tau)$.
When replacing, one should avoid possible missing or
overcounting of imbedded surfaces.  This is subtle point and it
will be considered in Section~5.  {\bf Let us emphasize that
this would be not a standard string theory}.

Let us also notice the case $D=4$ is special in
such an approach, since in this case the number of Euclidean
coordinates (three)  coincides with the number of components
of a two-dimensional metric. Because of that, in four-dimensional
case the number of dynamical fields (over which
goes integration in the path integral) remains unchanged.
Also for that case under certain conditions(see Section~5) the
$X^i(\sigma,\tau)$ can be restored unambiguously  from
$\glmn(\sigma,\tau)$ and, hence, there is a possibility to
define {\bf four-dimensional } correlational functions in the
theory (see Section~5).

Although this approach could seem artificial, it could
provide us with a possibility of the calculation of
correlational functions of a  four-dimensional
string theory using the technology of matrix models and
integrable theories. Actually, this is the ultimate goal
of this consideration.

\newsec{Action}

Let us consider the following nonlinear functional of
$g^{\mu\nu}(z)$($\mu,\nu=1,2$) and $\phi (z)$ where $z=(z^1,z^2)$
are coordinates on some two-dimensional
manifold $\Sigma_2$:
\eqn\es{ S[\glmn,\phi]=\int d^2z \sqrt{g}
\sqrt{g^{\mu\nu}\partial_\mu \phi \partial_\nu \phi -1}}

as an action of
two-dimensional field theory with gravitational field.
If one want to consider only time-like surfaces
in the framework of four-dimensional interpretation of \es\
(see Section 5),
than the domain
of definition of the metric $\glmn(z)$ and of the one scalar
field $\phi(z)$ should be defined by the condition of reality
of the action $S$, i.e.  \eqn\edd{ F\equiv g({\squ1})>0}
This condition is fully analogous to the usual condition of the
reality of the Nambu-Goto action:
\eqn\eddng{
(\dot X X^\prime)^2-\dot X^2 X^{\prime 2}>0
}

However, in the recent work of Carlini and Greensite \ref\rcarl{
A.Carlini and J.Greensite,\ {\it Preprint NORDITA-94/71P\/}
(gr-qc/9502023)} it was argued that in case of square-root
actions (like \es\ ) one should in path integral integrate
over time-like {\bf and} space-like trajectories (or surfaces)
in order to obtain unitarity and finiteness of the theory.
So, the question: should one impose the condition \edd\ ,
is still open (at least for the author).

Let us note that the action \es\  can be rewritten as follows
\eqn\esd{ S[\glmn,\phi]=\int d^2z
\sqrt{\det(g_{\mu\nu}-\partial_\mu \phi \partial_\nu \phi)}
}

If one consider $\glmn(z)$ as a dynamical variable than
the theory, defined by the action \es\ , is not a standard one
in a sense that its equations of motion (for $\ghmn$ or $\glmn$)
are contradictory (if there is only {\it one\/} field $\phi$).
The same problem (absence of
equations of motion for a metric or degeneracy of a metric)
actually arises in
{\bf any} two-dimensional theory of {\bf one} scalar
field coupled to gravity. The reason is very simple: the
equations of motion for the metric $\ghmn$ in such theories
always look like:
$$
A(z) \partial_\mu\phi\partial_\nu\phi + B(z) \glmn=0
$$
and, hence,
$$
\det (\glmn)=\det
(-\frac{A(z)}{B(z)}\partial_\mu\phi\partial_\nu\phi)\equiv 0
$$

The familiar examples are
Polyakov action \ref\rpol{ A.M.Polyakov,\ {\it Phys.Lett.\/}\
{\bf 103B} (1981) 207,211}\ with one-dimensional target space,
which eventually produces 2-dimensional critical string theory
with additional Lioville field (see \ref\rddk{ F.David,\
{\it Mod.Phys.Lett.\/}\ {\bf A3} (1988) 1651 \semi J.Distler and
H.Kawai,\ {\it Nucl.Phys.\/}\ {\bf B321} (1989) 509}\ ):  $$
S_{Polyakov}=\frac{1}{4\pi\alpha^\prime}\int d^2z \sqrt{g}
(\ghmn \partial_\mu\phi \partial_\nu\phi + \Lambda) $$

and Lioville action  \rgins\  :
$$
S_{Lioville}=\frac{1}{8\pi}\int d^2z \sqrt{g} (
\ghmn \partial_\mu\phi \partial_\nu\phi
+Q\phi R(g) + \frac{\mu}{\gamma^2} e^{\gamma \phi})
$$
\bigskip

However if we consider $\glmn$ as an external field and
not a dynamical variable, the equation of motion for
the dynamical field $\phi(z)$ exists:
\eqn\eem{
\Phi(\glmn(z),\phi(z))\equiv
\frac{1}{\sqrt{\det(g)}}
\frac{\delta S[\glmn,\phi]}{\delta \phi(z)}
=D^\rho\left(\frac{\partial_\rho\phi}{\sqrt{\squ1}}\right)=0
}

We can also calculate the energy-momentum tensor of the theory
\es\ using the standard definition:
\eqnn\etem
$$\eqalignno{
T_{\mu\nu}(\glrs(z),\phi(z))&=\frac{1}{\sqrt{g}}
\frac{\delta S[\glrs,\phi]}{\delta \ghmn(z)}  &\etem \cr
&=\frac{1/2}{\sqrt{\sq2}}
(\partial_\mu\phi\partial_\nu\phi-\glmn(\sq2))
\cr }
$$

It turns out that the action $S$ has some intriguing properties,
namely:

{\bf 1.} The action $S$ besides the two-parametric reparametrization
invariance has else one infinite-dimensional local symmetry.
All these symmetries allow to fix a gauge (locally) so that the
resulting theory becomes the theory of a free scalar massless
field. That new symmetry has a clear geometrical meaning.

{\bf 2.} The action $S$ equals to the area of the two-
dimensional manifold $\Sigma_2$ imbedded into four-dimensional
Mincowski space whereupon this imbedding is given
by $\glmn$ and $\phi$. Hence, when constructing a quantum theory
of random surfaces in four dimensions, one may try to replace the
path integration over $X^\alpha(z)$ ($\alpha=0,...,3$)
which usually describe an imbedding with the path
integration over $\glmn(z)$ and $\phi(z)$ and, expressing
$X^\alpha(z)$ via $\glmn(z)$ and $\phi(z)$, define
four-dimensional correlational functions.

{\bf 3.} The theory defined by $S$ has no anomalies, at least
naively (i.e. the symmetries of the action are preserved in
regularized theory).

\newsec{Symmetries}

The action $S$ is evidently
invariant under coordinate transformations :
$$
S^\prime\equiv S[g^\prime_{\mu\nu},\phi^\prime]=S[\glmn,\phi]
$$
where

$$\eqalignno{
z^\prime&=z^\prime(z)       \cr
\phi^\prime(z^\prime)&=\phi(z) \cr
g^\prime_{\mu\nu}(z^\prime)&=\frac{\partial z^\rho}
{\partial z^{\prime\mu}} \frac{\partial z^\sigma}{\partial
z^{\prime\nu}} g_{\rho \sigma}(z)  \cr}
$$

This invariance is equivalent to the following identity
for the energy-momentum tensor introduced  in Section~2:
\eqn\eidd{
D_\mu T^{\mu\nu}\equiv 0
}
where $D_\mu$ is the covariant derivative.

Furthermore, there is else one new infinite-dimensional symmetry
of the action (which we will further call "form-symmetry"):  $$
S^\zeta\equiv S[g^\zeta_{\mu\nu},\phi^\zeta]=S[\glmn,\phi] $$
where the "form-transformations" are

\eqna\eft
$$\eqalignno{
g^\zeta_{\mu\nu}(z)&=\glmn(z)+\partial_{(\mu}\phi\partial_{\nu)}\zeta
+\partial_\mu\zeta\partial_\nu\zeta &\eft a \cr
\phi^\zeta(z)&=\phi(z)+\zeta(z) &\eft b \cr
}
$$
where $\zeta(z)$ is an {\it arbitrary\/} function of $z$.

The invarariance of the action \es\ with respect to
form-transformations is equivalent to the following identity
between the energy momentum tensor $T^{\mu\nu}$ and the $\Phi$
introduced in Section~2:
\eqn\eidft{
\Phi\equiv T^{\mu\nu}D_\mu\partial_\nu\phi }
where identity \eidd\ was taken into account.

One can also mention that the generalization of the action \esd\
\eqn\esdgen{ S[\glmn,\phi^i]=\int d^2z
\sqrt{\det(g_{\mu\nu}-\partial_\mu \phi_i \partial_\nu \phi_i)}
}
where $\phi_i(z)$ ($i=1,..D$) are scalar fields, is also
form-symmetric if $D>1$, i.e. symmetric under the following
transformations:
\eqna\eftgen
$$\eqalignno{
g^\zeta_{\mu\nu}(z)&=\glmn(z)+\partial_{(\mu}\phi_i\partial_{\nu)}
\zeta_i+\partial_\mu\zeta_i\partial_\nu\zeta_i &\eftgen a \cr
\phi_i^\zeta(z)&=\phi_i(z)+\zeta_i(z) &\eftgen b \cr
}
$$
where $\zeta_i(z)$ are arbitrary functions of $z$.
The action \esdgen\ has no equations of motion for
metric $\glmn$ as well.

At the same time the possible generalization
of the action in the form \es\ :
\eqn\esgen{
S[\glmn,\phi^i]=\int d^2z \sqrt{g}
\sqrt{g^{\mu\nu}\partial_\mu \phi_i \partial_\nu \phi_i -1}
}
where $\phi^i(z)$ ($i=1,..D$) are scalar fields, is {\bf not}
form-symmetric if $D>1$.

Let us also notice that form-transformations are abelian and
have the summation rule $\zeta_{12}=\zeta_1+\zeta_2$, i.e.
$$\eqalignno{
(\glmnp{\zeta_1})^{\zeta_2}&=(\glmnp{\zeta_2})^{\zeta_1}=\glmnp{
\zeta_1+\zeta_2} \cr
(\phi^{\zeta_1})^{\zeta_2}&=(\phi^{\zeta_2})^{\zeta_1}=\phi^{
\zeta_1+\zeta_2} \cr }
$$

{\bf Remark 1.} The form-transformations leave invariant not
only the action \es \ , but also the integrand \eqn\eddi{
F=g({\squ1})=inv }

Generically there are
transformations, which change the signature of the metric.
This is possible in particular because the action \es\
can be rewritten in the form \esd\ , where the metric
$\glmn$ appears {\it with only lower indices\/}, namely,
\eqn\esli{ S=\int d^2 z F^{1/2}} where \eqn\ed{
F=g_{00}(\partial_1\phi)^2- 2
g_{01}\partial_0\phi\partial_1\phi+ g_{11}(\partial_0\phi)^2-
  g_{00}g_{11}+g^2_{01}}

Hence, there is no singularity in the action in the point
$g=0$. Nevertheless, it is useful to find the
form-transformations that produce the degenerate metric.
For a given field configuration $\phi(z)$ and $\glmn(z)$
this degenerating transformations $\zeta(z)$ are given
by the following equation:
\eqn\edft{
h_{00}(\partial_0\zeta-\partial_0\phi)^2+
h_{11}(\partial_1\zeta-\partial_1\phi)^2-2
h_{01}(\partial_0\zeta-\partial_0\phi)(\partial_1\zeta-\partial_1\phi)
-F=0}
where
$$\eqalignno{
h_{00}&=g_{00}-(\partial_0\phi)^2 \cr
h_{11}&=g_{11}-(\partial_1\phi)^2 \cr
h_{01}&=g_{01}-(\partial_0\phi)(\partial_1\phi) \cr
}$$

It is important to note that any given metric can be transformed
(at least locally) via form-transformations to a metric
with a definite signature\footnote{$^1$}
{\baselineskip 12pt It is interesting that from \edft\ follows
that if we consider not a "time-like action" (i.e. real on
time-like surfaces, see Section 5) but a "space-like action"
\eqn\essl{ S[\glmn,\phi]=\int d^2z
\sqrt{g} \sqrt{1-g^{\mu\nu}\partial_\mu \phi \partial_\nu
\phi} }
which is real on space-like surfaces,
than there would be no form-transformation which makes the
metric degenerate
because in that case the equation \edft\ would have no
solution.} (say, positive) since for the fields belonging to the
domain of definition \edd\ the equation \edft\ always define
some hyperboloid on a plane $(\partial_0\zeta,\partial_1 \zeta)$
which divides the plane into the regions with different
signatures of the metric.

{\bf Remark 2.} The action \es\ is form-invariant also in case
when the dimension of the world-sheet  is not equal to two.

\newsec{Gauge fixing, resulting free action, tracelessness
of the energy-momentum tensor in the gauge}

The theory, defined by the action \es, is a gauge
theory with 3-parametric family of gauge transformations (2
parameters define diffeomorfism and 1 parameter defines
form-transformation). Now we can fix a gauge in the classical
theory in the following way (imposing 3 gauge condition):

\eqna\egc
$$\eqalignno{
g_{00}+g_{11}-(\partial_0 \phi)^2-(\partial_1 \phi)^2&=0
&\egc a \cr
g_{01}-\partial_0\phi\partial_1\phi&=0  &\egc b \cr
g_{00}-g_{11}&=0  &\egc c \cr }
$$

Let us also note that first two of these gauge conditions
(i.~e.~\egc{a,b}) are invariant with respect to
form-transformations and, hence, could be considered as
fixing only the reparametrization invariance. Furthermore,
in Section 5 below we will see that these two conditions
in the framework of a four-dimensional string interpretation
of the action \es\ come from the usual orthonormality
gauge  conditions in the Nambu-Goto string formalism
(see \ref\rng{ J.Scherk,\ {\it Rev.Mod.Phys.\/}\ {\bf 47} (1975)
123}\ ):  $$\eqalignno{ \dot X^2+X^{\prime 2}&=0 \cr (\dot X
X^\prime)&=0 } $$

By analogy with that case, we will call the first two conditions
\egc{a,b}\ "orthonormality conditions".

The third gauge condition \egc{c}\ is not form-invariant and
fixes the form-symmetry. This gauge condition can be always
resolved locally.

Now all the components of the metric can be written in terms of
$\phi$:
$$
(\glmn)=\frac{1}{2}\pmatrix{
           (\partial_0 \phi)^2+(\partial_1 \phi)^2 &
           2\partial_0\phi\partial_1\phi \cr
           2\partial_0\phi\partial_1\phi &
           (\partial_0 \phi)^2+(\partial_1 \phi)^2 \cr}
$$
and
$$
(\ghmn)=\frac{2}{((\partial_0 \phi)^2-(\partial_1 \phi)^2)^2}
           \pmatrix{
           (\partial_0 \phi)^2+(\partial_1 \phi)^2 &
           -2\partial_0\phi\partial_1\phi \cr
           -2\partial_0\phi\partial_1\phi &
           (\partial_0 \phi)^2+(\partial_1 \phi)^2 \cr}
$$

Thus, this 3 gauge conditions allow us to exclude completely the
metric from the action (in a classical theory). The field $\phi$
remains the only dynamical variable. Although the possibility of
substitution of gauge conditions into the action in general case
is a subtle point even on a classical level, we assume that in
this case the substitution is valid.

After simple calculations we find that the resulting action
is the action of free scalar massless field:
\eqn\esg{
S^G[\phi]=\pm\frac{1}{2}\int
d^2z((\partial_0\phi)^2-(\partial_1\phi)^2)=
\int d^2z \sqrt{\det(g^G(\phi))} }

The index  $G$ in $S^G$ and $g^G$ denotes that these terms are
taken in the gauge \egc{a,b,c}\ .  The indeterminacy of the sign
of the r.~h.~s. of \esg\ corresponds to indeterminacy of the
choice of zeroth and first coordinate.  Let us assume the $(+)$
sign.  Than the zeroth coordinate $z^0$ plays the role of the
time coordinate.

The equation of motion for the field $\phi(z)$ which follows
from the action \esg\ is
$$
\ddot\phi-\phi^{\prime\prime}=0
$$
and can also be obtained by
substituting of the gauge conditions \egc{a,b,c}\ into the
equation \eem\ .

In the particular gauge \egc\ we find
\eqna\etemg
$$\eqalignno{
T^G_{00}=-T^G_{11}&=\frac{1}{4}((\partial_0\phi)^2-(\partial_1\phi)^2)
&\etemg a \cr
T^G_{01}&=0 &\etemg b \cr }
$$
and, hence,
\eqn\etemt{
T^{G \mu}_\mu=\ghmnp{G}T^G_{\mu\nu}=0
}
i.e. in this gauge the energy-momentum tensor is traceless.

{\bf Remark 1.}
The choice of the third gauge condition as $g_{00}=g_{11}$
(together with orthonormality conditions) is equivalent to the
condition $T^{G \mu}_\mu=0$ and can be replaced by it. In other
words, our gauge is fixed by the orthonormality conditions and by
the condition of tracelessness of the resulting energy-momentum
tensor.

Let us stress that there is an ambiguity in the choice of
the third gauge condition \egc{c}\ and, moreover, still it is
unclear for us should one fix the whole form-symmetry or
should one fix only it's subgroup.

Apparently, one should
choose the gauge condition  in accordance with a topology of the given
two-dimensional manifold. The gauge condition \egc{c}\ can be
imposed locally,
but for imposing a global gauge one need to perform more
sophisticated analysis.

{\bf Remark 2.} One should be cautious about using in this theory
the term "gauge symmetry" and "gauge conditions", since in the
usual sence they refers to symmetries of the classsical
equations of motion and these are absent in the theory. We use
the term "gauge symmetry" in a sense that {\bf action} (and
not equations of motion) is invariant with respect to
transformations.

{\bf Remark 3.} One can try to impose the following
gauge condition: $\phi^\zeta(z)\equiv 0$ thus excluding the
field $\phi(z)$ from the theory. However, it can be argued that it is
actually a "bad" choice of a gauge condition for the reasons
which are discussed in Section 5.

\newsec{Interpretation as 4-dimensional
string theory and possible 4-dimensional quantization}

The action (functional) \es\ depending on $\phi(z)$ and
$\glmn(z)$  which has a positive signature\footnote{$^1$}
{\baselineskip 12pt Any metric can be transformed via
form-transformations to a metric having positive signature, see
Section 3.}
actually equals to the area of some definite 2-dimensional
manifold $\Sigma^2$ imbedded into the 4-dimensional Minkovsi
space $R^{3,1}$. The corresponding imbedding $$
I^{g,\phi}:  \Sigma^2\longrightarrow R^{3,1} $$ is given by
$\glmn(z)$ and $\phi(z)$ and is built as follows :

{\bf Step 1.} Let us find some {\bf isometric imbedding}
$I_0^g$ of a 2-dimensional surface $\Sigma^2$ into
3-dimensional Euclidean space $R^3$. $I_0^g$ realizes the metric
$\glmn$ as the metric induced on $\Sigma^2$ by the imbedding:
$$
I^*_0: \eta_{ij}^{(Euclidean)}\longrightarrow\glmn
$$

The imbedding $I_0^g$ {\it naively\/} exists and is unique
(modulo global shifts and rotations of the resulting surface in
Euclidean space ) because the number of the components of metric
(three:  $g_{00}(z^1,z^2), g_{01}(z^1,z^2),g_{11}(z^1,z^2)$)
coincides with the number of functions $X^i(z^1,z^2)$
$(i=1,2,3)$ which parametrically define the imbedding. In order
to prove the existence and uniqueness of such imbedding one has to
prove the
global existence and uniqueness of the solution of the nonlinear
nonhomogenous
differential equation
\eqn\edi{ \partial_\mu X^i \partial_\nu X^i=\glmn  } with
respect to $X^i(z)$.

The answer to this question strongly depends on the class of
imbeddings.
As it was shown in \ref\rgrom{Gromov M.L. and Rokhlin V.A.\
{\it Uspekhi Mat. Nauk\/}\ {\bf 25} (1970) n.5}, if one consider
imbeddings of the class $C^\infty$ (i.e. functions $X^i(z)$
belong to $C^\infty$) than any compact two-dimensional Riemann
manifold can be isometrically embedded into $R^{10}$. In other
words, in general case one
needs {\bf ten} smooth functions $X^i(z)$ in order to
realize an arbitrary metric $\glmn(z)$ by \edi\ .

However the situation considerably changes for the case
of $C^1$ imbeddings. Nash and Kuiper
\ref\rnash{Nash J.\ {\it Ann. Math.\/}\ {\bf 60} (1954) 383-396
\semi Kuiper N.\ {\it I.Proc.Konikl.nederl.acad.wetensch.\/}\ {\bf
A 58} (1955) 545-556}
have shown that if one consider
imbeddings of class $C^1$ than any closed two-dimensional
Riemann manifold can be isometrically imbedded into $R^3$!
This seems surprising:
for instance, there exists an isometric imbedding of the flat
torus into three-dimensional Eucledian space.
Evidently, for such imbeddings the
metric \edi\ belongs to the class $C^0$.

Uniqueness of such an imbedding is not clear,
but let us {\it assume\/}  that this imbedding is
unique.

{\bf Step 2.} Let us assign
\eqn\efc{
X^0(z)=\phi(z) }
where $X^0$ is the forth coordinate. Now we have one-to-one
(due to the assumption above)
correspondence:
\eqn\egx
{ (\glmn(z),\phi(z)) \Rightarrow
(X^0(z),X^1(z),X^2(z),X^3(z))
}
where four functions $X^\alpha(z)$
$(\alpha=0,...3)$ belong to $C^1$ and define parametrically some
imbedding of the $\Sigma^2$ into $R^{3,1}$.  We will call this
imbedding $I^{g,\phi}$.

The inverse to \egx\ mapping
\eqn\exg{
(\glmn(z),\phi(z)) \Leftarrow
(X^0(z),X^1(z),X^2(z),X^3(z))
}
is built simply by equating: $
\glmn=\partial_\mu X^i \partial_\nu X^i$ ($i=1,2,3$) and
$X^0=\phi$.

{\bf Preposition.} The action \es\ gives the area of the
2-dimensional surface $\Sigma^2$ imbedded into $R^{1,3}$ by the
imbedding $I^{g,\phi}$:
\eqn\esisa{
S[\glmn,\phi]=Area[\Sigma^2]_{I^{g,\phi}} }

{\it Proof.} This preposition is proved simply by substituting
of the \edi\ and \efc\ into the action \es. The resulting
expression is the area of the surface $\Sigma^2$ imbedded into
$R^{1,3}$, i.e.

$$\eqalignno{
S[\glmn\rightarrow\partial_\mu X^i \partial_\nu X^i,
  \phi\rightarrow X^0]&=
\int d^2z\sqrt{\det(-\eta_{\alpha\beta}\partial_\mu X^\alpha
\partial_\nu X^\beta)} \cr
&=\int d^2z\sqrt{\det( (\dot X X^\prime)^2-\dot X^2 X^{\prime
2})}
\cr }
$$
where $\eta_{\alpha\beta}=diag(+,-,-,-)$ is a Mincowski
metric and $\alpha,\beta=0,...,3$.

{\it Q.E.D.}

Thus, one may say that the theory \es\ defines some string
theory in 4 dimensions.

The metric induced on the $\Sigma_2$ by the imbedding
into four-dimensional Mincowski space is
\eqn\eim{
\tilde g_{\mu\nu}=g_{\mu\nu}-\partial_\mu \phi \partial_\nu \phi}

Let us note that in the gauge \egc{a,b,c}
$\det \tilde g=\det g$ (and this condition can replace the third
gauge condition \egc{c} fixing the form-invariance)

It is now easy to see that, in fact, first two of the three
gauge conditions \egc{a,b,c}, in the framework of the
4-dimensional interpretation, are equivalent to the following
gauge conditions usually imposed in canonical quantization of
the Nambu-Goto string theory:  $$\eqalignno{ \dot X^2+X^{\prime
2}&=0 \cr (\dot X X^\prime)&=0 } $$

\smallskip
{\bf Remark 1.}
One can write down the point particle action analogous to
the action \es\ :
\eqn\esp{
S_P[e(\tau),\phi(\tau)]=\int
d\tau\sqrt{e}\sqrt{e^{-1}\dot\phi^2-1} }
which is reparametrization invariant (
$\tau\rightarrow\tau^\prime(\tau)$) and form-invariant
($\delta_\zeta\phi=\zeta$, $\delta_\zeta e=2\dot\phi\dot\zeta$).
This action has the analogous geometrical meaning:  it
equals to the length of a corresponding curve in $1+1$-
Mincowski  space. This can be seen by substituting into the
action \esp\ $e=\dot\varphi^2$. In fact, this point action is
useful as a some kind of a toy model for string action \es\ .

\smallskip

Let us stress the speciality of 4 dimensions for action \es\ . In
principle, we could consider the action \es\ as the area of the
surface imbedded into target Mincowski space which is
$D$-dimensional ($D>4$):  we would only need to realize
the metric $\glmn$ as induced by an imbedding of the surface
into $R^{D-1}$ ($D-1>3$) and this is certainly possible. But
than the correspondence between metric and imbedding
definitely would be not one-to-one.

Now having that interpretation of the action \es\ as the area of
the surface imbedded into 4-dimensional Mincowski space
one can try to develop some corresponding 4-dimensional string
theory. Let us {\bf define} correlational functions of that
hypothetic theory as follows:
\eqn\edcf{
\langle X^\alpha(z_1)...X^\gamma(z_n)\rangle_\theta=
\int D\glmn(z) D\phi(z)
X^\alpha_{g,\phi}(z_1)...X^\gamma_{g,\phi}(z_n) e^{-\theta S} }
where we denote
by $X^\alpha_{g,\phi}(z)$ the solution of equations \edi\ and
\efc\ ,i.e. {\bf they are expressed} via $\glmn(z)$ and
$\phi(z)$.  Integration goes over $\glmn(z)$ and $\phi(z)$ and
one should build a proper functional measure and indicate the
class of functions over which goes integration.  This is, in
fact, difficult problem which should be addressed elsewhere
and we, by analogy with  path integral of quantum mechanics,
just assume that integration goes over $\glmn(z)\in C^0$ and
$\phi(z)\in C^1$.  Than, due to above mentioned results of Nash
and Kuiper, in the path integral all imbeddings of Riemann
surfaces in four-dimensional Mincowski space are produced
without missing any surface and without overcounting any
surface.

Let us admit as $D_{\hat g}\glmn D_{\hat g}\phi$ the
standard formally defined Polyakov measure \rpol\  :  $ \int
D_{\hat g}\delta\phi\ \exp(-\int\sqrt{\hat g}(\delta\phi)^2)=1
$ and $ \int D_{\hat g}\delta\glmn\ \exp(-\frac{1}{2}
\int\sqrt{\hat g}(2\hat\glmn \hat g_{\rho\sigma}+ \hat
g_{\mu\rho}\hat g_{\nu\sigma})\delta\ghmn\delta
g^{\rho\sigma})=1 $

This functional measure is invariant with respect to
diffeomorfisms, i.e.
$D_{\hat g^\prime}\glmn D_{\hat g^\prime}\phi=D_{\hat g}\glmn
D_{\hat g}\phi$ (and, hence, $Z_\theta[\hat \glmnp{\prime}]=
Z_\theta[\hat \glmn]$). One can write equivalently:
\eqn\eidda{
\langle D_\mu T^{\mu\nu}\rangle=0 }
i.e. the vacuum amplitude of the classical identity \eidd
corresponding to diffeomorfism invariance equals to zero.

Now we should answer the question: is the measure
invariant with respect to form-transformations.
The possible anomaly of the measure can be obtained via
perturbation theory (see for a review
\ref\ralv{ L.Alvarez-Gaum\'e, in {\it Unified String Theories},
           World Scientific, 1986}\ ). In this approach
the presence of an anomaly is connected with violation of the
corresponding symmetry by the regulirization procedure. For
instance, Weyl invariance of the Polyakov string is violated by
the regularization (e.g. dimensional) and, hence, Weyl anomaly
arises.

In our case, since the form-symmetry is valid for all dimensions
of a world-sheet space (see the Remark~2 in Section~3),
dimensional regularization will not violate the form-symmetry
and there will be no "form-anomalies" in the theory. The vacuum
amplitude of the corresponding identity \eidft\ equals to zero
(if we use dimensional regularization):
\eqn\eidfta{
\langle\Phi\rangle=\langle T^{\mu\nu}D_\mu\partial_\nu\phi\rangle
}

{\bf Lorentz invariance}.
The correlational functions of the four-dimensional theory
defined by \edcf\ are globally Lorentz covariant. First, let us
show the covariance with respect to global Lorentz boosts.  In
order to show that let us consider two field configurations
(i.~e.~imbeddings) $X^\alpha(z)$ and $X^{\alpha\prime}(z)$
connected by an infinitesimal boost:  $$\eqalignno{
X^{0\prime}&=X^0+\epsilon X^1 \cr
X^{1\prime}&=X^1-\epsilon X^0 \cr
X^{2\prime}&=X^2 \cr
X^{3\prime}&=X^3 \cr
}$$

It is easy to see that two field configurations $(\glmn,\phi)$
and $(\glmn^\prime,\phi^\prime)$ corresponding to
$X^\alpha$ and $X^{\alpha\prime}$, accordingly, (see \exg\ )
are connected by an infinitesimal form transformation with
parameter $\zeta(z)=\epsilon X^1(z)$:
$$\eqalignno{
\glmn^\prime&=\glmn^\zeta |_{\zeta=\epsilon X^1} \cr
\phi^\prime&=\phi^\zeta |_{\zeta=\epsilon X^1} \cr
}$$
and, since the action $S$ is form-invariant an the measure is
equal to $e^{-S}$, both
configurations have in the path integral \edcf\ equal
weights.  Hence, the measure of the path integral is
invariant with respect to global Lorentz boosts. Let us also
notice that Lorentz transformations are realized on fields
$(\glmn,\phi)$ {\it nonlocally}.

Since the mappings \egx\ and \exg\ are invariant with respect
to global $SO(3)$ rotations in the space $X^i$ ($i=1,2,3$),
the measure of the funcional integral \edcf\ is globally
$SO(3)$-invariant as well. Together with global boost
invariance of the measure this yields the global Lorentz
covariance of the correlational functions, defined by \edcf\ .

Here comes some subtle point. If we are going to consider
form-symmetry as a gauge symmetry than we have to define
as a physical variables only scalars of a gauge group.
Since the Lorenz transformations are subset of the gauge group,
one should consider only Lorentz {\it invariants\/} as physical
variables. But we want Lorentz {\it covariants\/} to be included
into the set of physical variables as well. There are two
way-outs:

Approach 1. Factorize not over the whole group of
form-transformations but over form-ransformations minus Lorentz
transformations.  In this case gauge conditions must be Lorenz
invariant.  Than one can use usual definition of physical
variables (i.e. scalars of the gauge group).  In this approach
one should demonstrate the invariance of the measure of
path integral with respect to (non-local) Lorentz
transformations. This invariance can depend on the choice of a
gauge condition.

Approach 2. Factorize over the whole group of form
transformations and consider as physical variables not only the
scalars of the gauge group but also the variables which
transform {\bf covariantly} under the action of Lorentz
transformations and {\bf invariantly} under the action of the
other form-transformations.  This does not coincide with the
standard definition of the physical variables.  In this approach
the vacuum amplitudes of physical variables {\it do depend\/} on
the choice of gauge condition in the following way: they
transform covariantly under Lorentz transformations of the gauge
condition.

In Section~4 (Remark~3) we noted that the gauge condition
$\phi(z)\equiv 0$ is not good. The argument for that comes from
Lorentz covariance. This condition evidently is not Lorentz
invariant and, hence, cannot be used in the first approach.
It is also not good for the second approach since problems
with definition of the physical variables arise.

Let us notice that the third gauge conditions \egc{c}\
in Chapter~3 is non-Lorentz invariant.

\smallskip

\newsec{Discussion}

The main goal of the consideration of the action \es\
is a possibility to apply {\bf matrix model technology} for
the {\bf four-dimensional string theory}.

It is well known
(see \rgins\ ) that matrix models correspond
to the case of (initially) {\bf one-dimensional} target space,
i.~e.~when there are only one field $X(z)$ (and
metric $\glmn(z)$ which eventually produces second field:
Lioville field $\varphi(z)$).

One can demonstrate this as follows (see \rgins\ for details).
Consider the partition finction
\eqn\epfst
{Z=\sum_{g}\int D\glmn(z) D\phi(z) e^{-S_0}}
where
\eqn\esl{S_0=\int d^2z \sqrt{g}(\frac{1}{\alpha^\prime}
\ghmn\partial_\mu X\partial_\nu X + \Phi R + \lambda)}

Discretizing the action \esl\ (approximating surfaces by
collection of equilateral triangles of area
$S_\bigtriangledown$) we get for path integral \epfst\
\eqn\epfste { Z(g_0,\kappa)=\sum_h g_0^{2h-2}\sum_\Lambda
\kappa^V\prod_{i=1} ^V\int dX_i\prod_{<ij>}e^{-(X_i-X_j)^2} }
where $g_0=e^\Phi$, $\kappa=e^{-\lambda S_\bigtriangledown}$.
First sum runs over genus of discretized surfaces. Second sum
runs over all distinct lattices $\Lambda$ and $V$ is a number of
triangles in a lattice. Second product in \epfste\ runs over
links of a dual lattice ($X_i$ live on the vertices of a dual
lattice). The term $e^{-(X_i-X_j)^2}$ arised at the rhs of
\epfste\ because the discretized version of $\int d^2z
\sqrt{g} \ghmn\partial_\mu X\partial_\nu X$ is simply
$\sum_{<ij>} (X_i-X_j)^2$, where the sum runs over all the links
of the dual lattice.

As was first noted by Kazakov and Migdal\ref\rkm{V.Kazakov and
A.Migdal,\ {\it Nucl.Phys.\/}\ {\bf B311} (1989) 171}, a
statistical sum of the form \epfste\ is generated in the Feinman
graph expansion of the quantum mechanics of a $N\times N$
hermitian matrix. Consider the Euclidean path integral
\eqn\epfmm
{Z=\int D^{N^2}\Phi(\tau) \exp \left[-\beta \int^T_Td\tau Tr(
\frac{1}{2}\dot \Phi^2 + \frac{1}{2\alpha^\prime}\Phi^2-
\frac{1}{6}\Phi^3)\right]}
One obtains the sum over all connected Feynmann graphs
$\Lambda$:
\eqn\epfmme
{\lim_{T\rightarrow \infty}\ln Z=\sum_h N^{2-2h}\sum_\Lambda
\kappa^V\prod_{i=1}^V\int_{-\infty}^\infty d\tau_i\prod_{<ij>}
e^{-|\tau_i-\tau_j|/\alpha^\prime}}
where $\kappa=\sqrt{N/\beta}$.
The exponential $e^{-|\tau_i-\tau_j|/\alpha^\prime}$
is the one dimensional massive Euclidean propagator
in configuration space.
The point is that this expression almost coinsides with
\epfste\ .

The only difference between the two expressions is the
exponentials:  for the case of two-dimensional gravity we have
$e^{-(X_i-X_j)^2}$ while for the case of matrix model we get
$e^{-|\tau_i-\tau_j|/\alpha^\prime}$. Usual point of view is that
the exact expessions for the exponential are not important
especially as matrix model calculations based on partition
function \epfmm\ coincide with those (few) results obtained from
continiuos calculations in two-dimensional gravity \esl\ .

Let us now consider the action \es\ . First, let us note that in
the action \es\ the field $\phi(z)$ must have the dimension of
length in order to the expression
$\ghmn\partial_\mu\phi\partial_\nu\phi$ be dimensionless. If one
wish to consider dimensionless field $\phi(z)$, one should add
some dimensional parameter $c$ ($[c]=[l]$) into the action:
\eqn\esc{
S_c[\glmn,\phi]=\int d^2z \sqrt{g}
\sqrt{c^2 g^{\mu\nu}\partial_\mu \phi \partial_\nu \phi -1}
}

Since $\phi$ plays the role of fourth coordinate $X^0$ (or time)
in the framework of the four-dimensional interpretation of the
action \es\ (see Section~5, Step~2), we can write:
$$
\ghmn\partial_\mu\phi\partial_\nu\phi \sim 1/{v^2}
$$
where $v$ is four-dimensional velocity of a point on the
propagating string. Hence the action \esc\ is
$$
S_c\sim \int d^2z\sqrt{\frac{c^2}{v^2}-1}
$$
and the physical sence of parameter $c$ is simply the speed of
light.

The case $\frac{c^2}{v^2}\rightarrow\infty$ corresponds to
non-relativistic consideration. In this limit we have
\eqn\escnr{
S_{NR}=\lim_{c\rightarrow\infty} S_c=c \int d^2z \sqrt{g}
\sqrt{g^{\mu\nu}\partial_\mu \phi \partial_\nu \phi } }

And discretizing this action analogously to \esl\
and considering the partition function we get
the expession \epfmme\ {\bf with correct
exponentials} $e^{-c|\tau_i-\tau_j|}$, where one can
identify $c=1/\alpha^\prime$.

Hence, considering action \esc\ and it's non-relativistic
limit one can obtain the {\bf exact} coincideness
of the partition functions of 2-dimensional gravity in the
form \esc\ and matrix quantum mechanics \epfmm\ .

{\bf Remark 1.} Considering another limit $c\rightarrow 0$
(actually $c^2/v^2\rightarrow 0$),
which corresponds to the case when all velocities are
much greater than speed of light, we get Polyakov action
with one dimensional target space:
$$
S_{c\rightarrow 0}=i\left[\int d^2z\sqrt{g}-c^2\int d^2z
\sqrt{g} g^{\mu\nu}\partial_\mu \phi \partial_\nu \phi\right]
$$
\bigskip

\bigbreak\bigskip\bigskip\centerline{{\bf Acknowledgements}}\nobreak
I would like to thank Professor I.V.Tyutin and Professor
M.~A.~Soloviev for very useful
and invaluable discussions.
The work was supported in part by
the Landau Scholarship (KFA, J\"ulich, Germany).

\bigskip

\appendix{A}{Geometrical meaning of the form-symmetry}

Geometrical meaning of the form-symmetry becomes clear in
the framework of the four-dimensional interpretation of the
action \es\ . This infinite-dimensional symmetry in fact describes
the transformation of the shape (or form) of a  two-dimensional
surface imbedded into four-dimensional target space without
change of it's area. For example, one can imagine a
two-dimensional sphere and, evidently, it can be {\bf
crumpled} having it's area (i.e. action \es\ ) fixed.
Intuitively, it is clear that this is indeed
infinite-dimensional symmetry.

It is important to note that this symmetry has nothing in
common with the reparametrization invariance which
maps the image of the imbedding onto itself and, hence,
does not change the shape of the imbedded surface.
In some sense, the reparametrization invariance is
an internal symmetry, while form-symmetry is an
external symmetry.

Also, one should distinguish the form symmetry from
the area-preserving diffeomorfisms of the target space,
because the latter transformations transform a
metric of the target space $G_{\alpha\beta}$
and the form-symmetry in our interpretation
does not transform it (the action \es\ is an area
of a surface in Mincowski space). In other words,
the form-symmetry does not affect on the target space but
changes only an imbedding.

Let us note that the Nambu-Goto action,
which is an area of the imbedded surface as well,
has also the form-symmetry. But for that case
it is not a {\it local\/} symmetry, i.e. the
form-transformations cannot
be written as local transformations of the
fields $X_\alpha(z)$ and, hence, are not gauge transformations.
The point is that if we consider $\glmn$ and $\phi$ as
dynamical variables of the theory (instead $X_\alpha$) the
form-symmetry becomes local in new variables and
becomes the real gauge symmetry.

\bigskip

\listrefs
\bye